\documentclass[conference]{IEEEtran}
\IEEEoverridecommandlockouts
\usepackage{cite}
\usepackage{amsmath,amssymb,amsfonts}
\usepackage{algorithmic}
\usepackage{graphicx}
\usepackage{textcomp}
\usepackage{xcolor}
\def\BibTeX{{\rm B\kern-.05em{\sc i\kern-.025em b}\kern-.08em
    T\kern-.1667em\lower.7ex\hbox{E}\kern-.125emX}}

\usepackage{amsmath}
\usepackage{multirow}
\usepackage{arydshln}
\usepackage{enumitem}
\usepackage{threeparttable}
\usepackage{graphicx}
\usepackage{caption}
\usepackage{subcaption}
\usepackage{ragged2e}
\usepackage{float}
\usepackage{algorithmic}
\usepackage{multicol}
\usepackage{caption}
\usepackage[ruled,vlined]{algorithm2e}
\newtheorem{definition}{\textbf{Definition}}[section]
\usepackage{makecell}
\usepackage{xcolor,colortbl}

\definecolor{LightCyan}{rgb}{0.8, 1.0, 1.0}
\definecolor{Gray}{gray}{0.9}

\definecolor{LightGreen}{rgb}{0.9, 1, 0.9}
\newcolumntype{a}{>{\columncolor{LightCyan}}c}
\newcolumntype{e}{>{\columncolor{Gray}}c}
\newcolumntype{d}{>{\columncolor{LightGreen}}c}

\begin{document}

\title{Event-based Product Carousel Recommendation with Query-Click Graph\\
}

% \and
% \and
% \and
% \and
% \and

\author{\IEEEauthorblockN{Luyi Ma, Nimesh Sinha, Parth Vajge, Jason H.D. Cho, Sushant Kumar, Kannan Achan}
\IEEEauthorblockA{\textit{Walmart Global Tech},
Sunnyvale, CA \\
{\{luyi.ma, nimesh.sinha, parth.vajge, jason.cho, sushant.kumar, kannan.achan\}}@walmart.com}
}

\maketitle

\begin{abstract}
% Many current recommender systems mainly focus on the product-to-product recommendations for the target product and user-to-product recommendations for the target user rather than modeling the typical recommendations for the target event (e.g., festivals, seasonal activities, or social activities) where multiple product carousels (i.e., lists of products with different event-related aspects) are recommended to address the multiple aspects of the shopping demands. 
Many current recommender systems mainly focus on the product-to-product recommendations  and user-to-product recommendations even during the time of events rather than modeling the typical recommendations for the target event (e.g., festivals, seasonal activities, or social activities) without addressing the multiple aspects of the shopping demands for the target event. 
Product recommendations for the multiple aspects of the target event are usually generated by human curators who manually identify the aspects and select a list of aspect-related products (i.e., product carousel) for each aspect as recommendations.
However, building a recommender system with machine learning is non-trivial due to the lack of both the ground truth of event-related aspects and the aspect-related products.
To fill this gap, we define the novel problem as the event-based product carousel recommendations in e-commerce and propose an effective recommender system based on the query-click bipartite graph. 
We apply the iterative clustering algorithm over the query-click bipartite graph and infer the event-related aspects by the clusters of queries. 
The aspect-related recommendations are powered by the \textit{click-through} rate of products regarding each aspect. We show through experiments that this approach effectively mines product carousels for the target event.
\end{abstract}


\section{Introduction}

Product recommendation tasks are essential for improving the user experience by recommending related products based on users' demands.
One of typical shopping demands is the demand for products related to an event (e.g., festivals like \textit{Christmas}, Sports events like the \textit{Super Bowl}, or seasonal activities like summertime outdoor activities).
Typically, an event refers to a planned occasion or activity. 
Without loss of generality, we refer to any festival, seasonal or social activity as an event in this paper.
Each event usually covers multiple aspects where users express their shopping demands. 
For example, \textit{Valentine's Day} might cover the aspects such as gift cards, candies, flowers and \textit{Super Bowl} might cover the aspects of parties, decorations and snacks. 
Thus, simply recommending the event-relevant products without distinguishing multiple aspects would hurt the user experience because users need more time to find items of an aspect which are scattered around in the product pool. 
An effective way to address each aspect of an event is via product carousels.
A product carousel contains a list of relevant items serving the same shopping demand. 
We can have several product carousels for an event to address multiple aspects separately.
% A useful way to address the shopping demand for event-related products is via product carousels where the general demand for the event is broken down into one or multiple related topics and each topic is associated with a list of topic-related products (i.e., a product carousel).
These product carousels are important because they summarize each event-related aspect for users by collecting the aspect-related products together in the same product carousel.
% Product carousels are important because they summarize the event-related topics for users by bring topic-related products together.
Figure 1 shows an example of four product carousels on Walmart.com homepage for \textit{Labor Day}.
We can see that \textit{Labor Day} are divided into four narrow shopping aspects in this example and each one is addressed by a product carousel where several items are collected for each individual aspect.

\begin{figure}[t]
    \centering
    \includegraphics[width=0.47\textwidth]{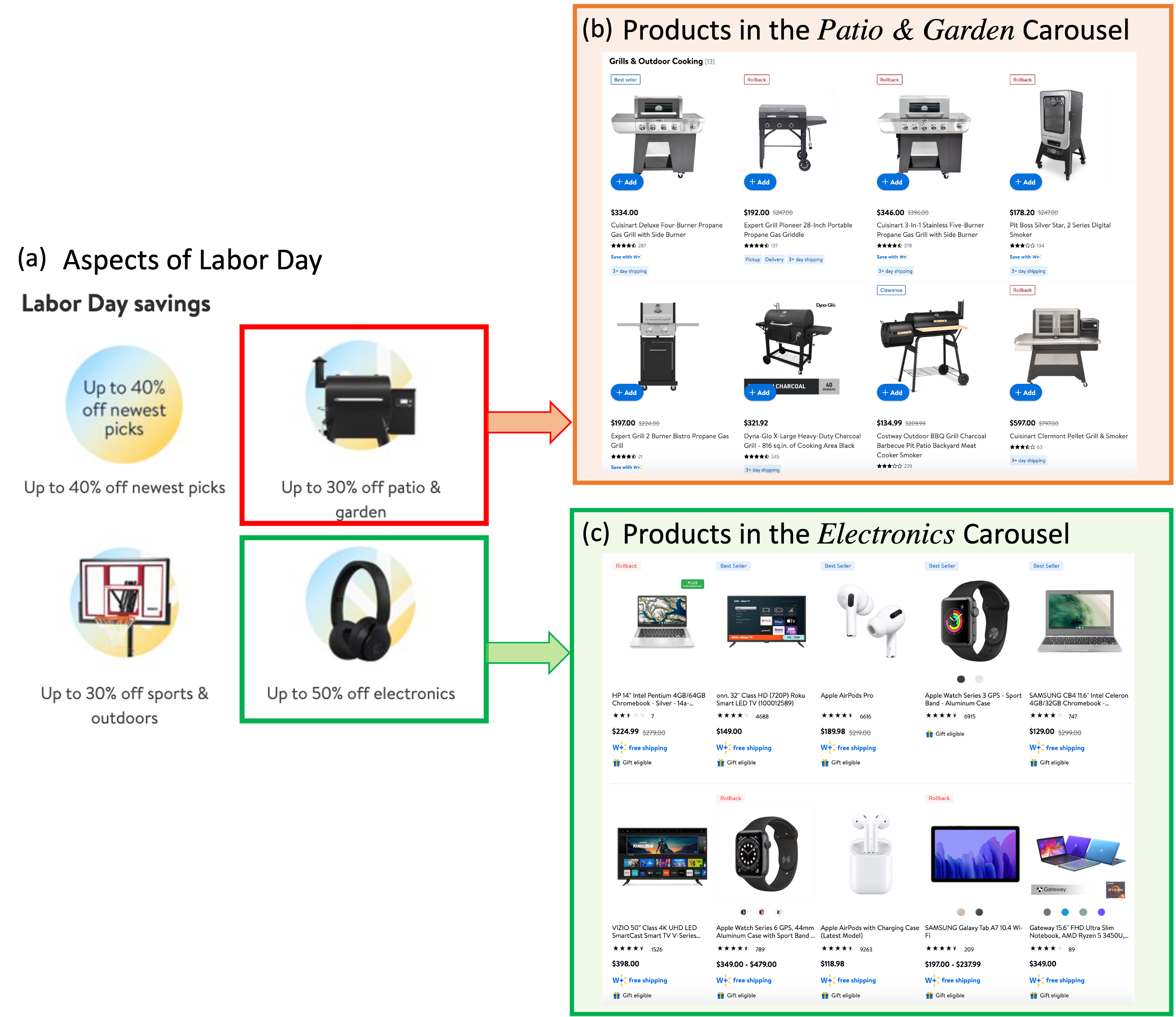}
    \caption{
    Product Carousels for 
    \textit{Labor Day} 
    on Walmart.com.
    Each carousel, presented by its icon in (a), represents a event-related aspect and  users can browse the product carousel of a certain aspect by a click on the carousel's icon, for example, the \textit{patio} carousel in (b) and the \textit{electronics} carousel in (c). 
    }
    \label{fig:product_carousel}
\end{figure}

% \noindent \textbf{paragraph 3 -- Requirement of Event-based product carousel recommendations and why current models fails: }
% However, many of current recommender system mainly focus on the user-to-item and item-to-item relationship, and it is hard to represent the carousel recommendations for event. 
However, building a recommender system that can automatically mine the meaningful aspects as well as their product carousels for an event is challenging due to the lack of ground truths of event-related aspects and the aspect-related products.
Moreover, the events and the product catalog evolve over time.
% However, automatically recommending meaningful product carousels by a recommender system with event-related topics for an event is challenging due to the lack of ground truths of event-related topics and the topic-related products because the events and the product catalog are evolving over time.
Hence, the content curation of both event-related aspects and their product carousels is usually completed by human curators
\footnote{https://www.theverge.com/21268972/hbo-max-design-recommendation-human-curation-friends-kids-profiles} who manually determine the aspects of the event and select aspect-related products from thousands of products to serve customers.
Many current recommender systems mainly focus on the product-to-product relatedness (e.g., substitutional products and complementary products \cite{DBLP:conf/cikm/LiuGDGGBY20}
\cite{DBLP:conf/kdd/McAuleyPL15}\cite{DBLP:conf/wsdm/WangJRTY18}) and user-to-product relatedness (e.g., personalized recommendations \cite{DBLP:conf/sigir/ChenYYH0020}\cite{ijcai2019-367}\cite{rendle2012bpr}\cite{DBLP:conf/wsdm/TangW18}).
They fail to recommend these event-related product carousels because of two reasons: 
(1) The product-to-product recommender systems take a product as the query and retrieve relevant products while the user-to-product recommender systems take a user as the query and generate product recommendations for this user. Both of them couldn't treat an event as the query and generate proper recommendations.
(2) Both the product-to-product and the user-to-product recommender systems couldn't automatically mine the multiple aspects of an event while generating the recommendations. 
% todo: this needs more explanation because the difficulty is not group the pt together, but the query, and it is required to form the product carousel.
% todo: show some examples of the existing RS, e.g., CF, MF, embedding-based. For example, they are either product-anchored, or user-anchored.
% todo: show the importance of events
% Although product search engines can link the event to related products by retrieving relevant products based on users' queries of event-related keywords, they still fail to generalize the topics of an event automatically.
Although product search engines can link the event to related products by retrieving relevant products based on event-related keywords in users' queries, they still fail to generalize the aspect of an event automatically.

% \noindent \textbf{paragraph 4 -- How to bridge the gap between the event and items, and automatically mine the topics of events for items: }
% introduce the data we use and our model.
% summarize our contribution.
% In this paper, we focus on the novel problem of event-based product carousel recommendations in e-commerce and propose an effective solution by bridging the gap between the target event and its topics of products via the search-click graph.
% In particular, we consider the user search behavior on the e-commerce platform as the source data and extract the event-related search queries and their clicked products for clustering.
% We consider the product-type \footnote{product type defines a set of products with the same attributes or functionalities. Products with the same product-type are substitutional to each other.} rather than products to keep the generality of the model and also reduce the sparsity.
% Following the success of query clustering in \cite{DBLP:conf/kdd/BeefermanB00}\cite{45569}\cite{DBLP:journals/tois/WenNZ02}, we conduct an iterative clustering method over the bipartite graph constructed by the event-related search queries and their clicked product-types to learn the topics of product carousels for an event.
% Each learned cluster of search queries can be viewed as a topic and the top-clicked product-types can be used to retrieve the actual products for the topic.
In this paper, we focus on the novel problem of the event-based product carousel recommendations in e-commerce and propose an effective recommender system for this problem by bridging the gap between the target event and relevant products of each event-related aspect via the query-click graph.
Typically, we construct the bipartite graph based on the event-related search queries and their click data to link the event-related aspects expressed in the queries and relevant product in the click data together.
To reduce the sparsity and increase the generality, we consider product-type \footnote{A product type defines a set of products with the same attributes or functionalities. Products with the same product-type are substitutional to each other. We consider product-types rather than products to keep the generality of the model and also reduce the sparsity. } rather than the individual product in the bipartite graph.
Following the success of the query clustering techniques, we conduct the iterative clustering method over the bipartite graph to group queries with similar aspects in the same cluster to mine aspects.
To build a recommender system for the event-based product carousel recommendations, we treat each learned cluster as a carousel and rank its associated product-types by the \textit{click-through} rate.
To evaluate the performance of our proposed solution, we compare our algorithm-generated carousels with the human-curated product-carousels and study the quality of learned clusters by computing the cohesion of recommended products in the same carousel and the heterogeneity between different carousels.

As a summary, the contributions of our paper are:
(1) We define the novel recommendation problem of the event-based product carousel recommendations 
(2) We propose an effective solution for building a recommender system of the event-based product carousel recommendations by conducting clustering over the query-click bipartite graph.
(3)  We evaluate the proposed solution by comparing the model-generated results with the ground truths labeled by human experts and show the effectiveness of our solution.
% For the rest of our paper, we summarize the literature review in Section \ref{sec:related_work} and present the modeling details in Section \ref{sec:model}.
% We report our experiment results in Section \ref{sec:experiment}.

\section{Related Work}
\label{sec:related_work}
\noindent \textbf{Bundle Recommendation}:
% collection, bundle, 
% https://dl.acm.org/doi/10.1145/3298689.3347003
% slate recommendation
The bundle recommendation problem (BRP) focuses on recommending a collection of items which serve a common aspect and could be co-purchased together. 
For example, Zhu et al. in \cite{DBLP:conf/sigir/ZhuHLT14} first study the BRP and reveal the effectiveness of solving the small-scale BRP.
Kouki et al. propose to incorporate  both domain knowledge from product suppliers as well as textual attributes from the products to learn the bundle recommendations in \cite{DBLP:conf/recsys/KoukiFVYAJQ19}.
However, our work is different from the BRP problem because the event-based product carousel recommendation problem not only requires the generation of aspect-related products but also need to determine a list of event-related aspects automatically, which is not solved by the BRP.

\noindent\textbf{Query Clustering}:
Many studies try to approach the query clustering problem.
Beeferman et al. first propose the agglomerative iterative clustering method in \cite{DBLP:conf/kdd/BeefermanB00} to cluster queries together based on the bipartite graph of queries and clicks from the search engine. 
Later, Wen et al. in \cite{DBLP:journals/tois/WenNZ02} use  the DBSCAN algorithm \cite{DBLP:conf/kdd/EsterKSX96} to cluster queries based on user logs.
Zhang et al. leverage the sequential search behavior of users in \cite{DBLP:journals/asc/ZhangN08} to group similar queries.
Radlinski et al. in \cite{DBLP:conf/www/RadlinskiSC10} consider the query reformulation data and clicks to cluster similar queries.
Recently, Kong et al. revisit the the agglomerative iterative clustering  in \cite{45569}  and use the estimated \textit{click-through}  rate as the edge weight. 
Unfortunately, they are not designed for product recommendations.
Our work extends the ideas of the iterative clustering method to the novel event-based product carousel recommendations with two new contributions: (1) we study the different roles of query data and the product-types in the iterative clustering under the product recommendation setting which is not addressed by  \cite{DBLP:conf/kdd/BeefermanB00}\cite{45569} and (2) we bridge the gap between query analysis and product carousel recommendations.

\section{Event-based Product Carousel Recommendation}
\label{sec:model}
In this section, we first establish the setting of our work by defining several important concepts.
Later, we describe the details of our event-based product carousel recommender system. 

\subsection{Problem Definition}
The event-based product carousel recommendation problem can be formally defined in Definition \ref{def:definition}. 
For example, let the event $E$ be `Valentine's Day'. the event-based product carousel recommender system needs to first infer $N$ aspects of `Valentine's Day' automatically such as `greeting cards', `stuffed toys', `candies' and so on. 
In addition, the recommender system needs to also provide recommendations for each aspect, which forms $N$ different product carousels to address multiple aspects of `Valentine's Day'.
\begin{definition}{\textbf{\textit{Event-based Product Carousel Recommendation}}:}
Given an event $E$ as the query, the recommender system needs to infer the $N$ potential aspects of $E$, $\mathcal{A}_E = \{A_1, A_2, ..., A_N\}$, and recommends a list of product (product-type) recommendations $R_{A_i} = \{r_1, r_2, ..., r_M\}$ as a product (product-type) carousel for each aspect where $A_i \in \mathcal{A}_E$.
\label{def:definition}
\end{definition}

With this definition, we can see its difference from the product-to-product recommendations and the user-to-product recommendations.
First, the event-based product carousel recommendations problem consumes the target event, which is different from the target product in the product-to-product recommendations and the target user in the user-to-product recommendations.
Second, existing product-to-product recommendation models mainly focus on the product-to-product relationship and fail to address the aspects of an event.
Although bundle recommendations \cite{DBLP:conf/sigir/ZhuHLT14}
\cite{DBLP:conf/recsys/KoukiFVYAJQ19} and basket-level complementary recommendations \cite{DBLP:journals/corr/abs-2010-11419}
\cite{DBLP:conf/sdm/LiuWGAY20} try to learn the common shopping aspects among products, these learned aspects are still associated with products instead of the target event.
Many user-to-product recommendations models try to personalize the recommendations, e.g., \cite{rendle2012bpr}\cite{DBLP:conf/cikm/WanWLBM18}\cite{DBLP:journals/isci/LiZLHYNJWZ21}.
However, none of them have the capability to model the event-related product carousels.

In the rest of the paper, we solve the event-based product carousel recommendation problem on the \textbf{product-type} level for generality because: (1) the product-level recommendations can be easily generated by the existing recommendation techniques (e.g., popular products, trending products with sale forecasting \cite{fan2017product}\cite{qi2019deep} or personalized product recommendations \cite{DBLP:conf/sigir/ChenYYH0020}\cite{ijcai2019-367}\cite{rendle2012bpr}\cite{DBLP:conf/wsdm/TangW18}) with products from the recommended product-types and (2) the product-level recommendations could be further optimized by downstream steps such as inventory requirements \cite{chen2020inventory} and the promotion of sponsored products.

\subsection{Iterative Query Clustering}
As aforementioned, the two key steps of modeling event-based product carousel recommendations are (1) inferring the aspects of the target events and (2) retrieving aspect-related product-types.
In this part, we build our recommender system for the event-based product carousel recommendaitons by organically combining the target event, aspects and product-types via the query-click bipartite graph with the iterative clustering algorithm.

\noindent \textbf{Search Queries and Click Data}:
The search query and click data naturally combine the users' shopping demands (by the user-provided keywords in queries) and its interaction with products (by users' clicks) together.
To enforce the relatedness between search queries and the users' shopping demands for the target event, we mainly consider the search queries which contains the keywords of the target event\footnote{Selecting the queries with keywords can be done by multiple ways, for example using Regular Expression. In our work, we first standardize the queries by converting each character into its lower case and apply the designed regular expression to select matched queries.}, as well as the corresponding impression\footnote{In our setting, the \textit{impression} refers to the number of times a product (product-types) retrieved by the search engine is presented to the users.} and click\footnote{In our setting, the \textit{click} refers to the number of times a product (product-type) is clicked by users.} data.
For example, for \textit{Valentine's Day}, we collect search queries with keywords like \texttt{valentines day} and \texttt{valentine day}.
Formally, let $\mathcal{Q} = \{q_1, q_2, ...\}$ denote the collection of all selected queries with the event-related keywords and $\mathcal{P}$ denote the collection of all the product-types in which each product-type $p \in \mathcal{P}$ has at least one impression record in the dataset.
For each search query $q \in \mathcal{Q}$, we denote the impression and click data of a product-type $p \in \mathcal{P}$ as $Imp_{(q, p)}$ and $Clk_{(q, p)}$ respectively.

\noindent\textbf{Bipartite Graph}:
% Only inferring the topics from either clusters of search queries or clusters from the product-site is not enough to bind the target events and topics of products together and also achieves less optimality \cite{DBLP:journals/tois/WenNZ02}.
% Hence, we consider the links between search queries and the product-types and construct a bipartite graph above the training data.
A bipartite graph $\mathcal{G}$ of search queries and product-types can be represented as $\mathcal{G} = \{\mathcal{V} = \{\mathcal{Q}, \mathcal{P}\}, \mathcal{E}\}$ where $\mathcal{V}$ represents the union of two independent vertex sets $\mathcal{Q}$ and $\mathcal{P}$ in the graph and $\mathcal{E}$ represents the set of edges between two types of vertices.
Each link between a pair of query-click tuple $(q, p)$, denoted as $e_{(q, p)} \in \mathcal{E}$,  represents that the product-type $p$ is clicked for the query $q$ and $w_{(q, p)}$ represents the edge weight.
Kong et al. in \cite{45569} proposed to estimate the edge weight via the Bayesian point estimator and we adapt this idea to our model for edge weights (Equation \ref{eq:edge_weight}).

\noindent\textbf{Iterative Clustering}:
Directly applying clustering algorithms on the query-click bipartite graph might not be optimal because similar queries might be apparently shown by the bipartite graph due to the sparsity of the query-click data \cite{DBLP:conf/kdd/BeefermanB00}.
To learn the product-type carousels with the event-related aspects, we extend the ideas of query clustering in \cite{DBLP:conf/kdd/BeefermanB00}\cite{45569} and mine the aspects as well as their associated product-types by iteratively clustering the query vertices in $\mathcal{Q}$ over the query-click bipartite graph $\mathcal{G}$.
In particular, we use the agglomerative hierarchical clustering as the base clustering method because the number of clusters is not available due to the lack of ground truths.
Let $a^{i} \in {A}_{\mathcal{Q}}^{i}$ and $b^{i} \in {B}_{\mathcal{P}}^{i}$ represent a cluster of query nodes and a cluster of product-type nodes after the $i$-th round of hierarchical clustering respectively, where both $a^{i}$ and $b^{i}$ are sets of query nodes and product-type nodes respectively.
When $i = 0$, each node forms a cluster by itself.
For the next round of clustering, we treat each cluster as a node and construct the feature vector by Equation \ref{eq:query_vector} \& \ref{eq:pt_vector} for each query cluster and product-type cluster respectively with the new edge weights computed by Equation \ref{eq:edge_weight} using the aggregated click and impression, $Imp_{(a^i, b^i)} = \sum_{q \in a^i, p \in b^i} Imp_{(q, p)} $ and $Clk_{(a^i, b^i)} = \sum_{q \in a^i, p \in b^i} Clk_{(q, p)} $.
% \begin{equation}
%     Imp_{(a^i, b^i)} = \sum_{q \in a^i, p \in b^i} Imp_{(q, p)} 
%     \label{eq:new_imp}
% \end{equation}
% \begin{equation}
%     Clk_{(a^i, b^i)} = \sum_{q \in a^i, p \in b^i} Clk_{(q, p)} 
%     \label{eq:new_clk}
% \end{equation}
% we construct the feature vector of each query cluster $q \in \mathcal{Q}$ (product-type $p \in \mathcal{P}$) based weights of the associated edges in $\mathcal{G}$.
% Formally, let $v_{q} \in \mathbb{R}^{|\mathcal{P}|}$ and $v_{p} \in \mathbb{R}^{|\mathcal{Q}|}$ denote the feature vector of a query $q$ and a product-type $p$ respectively, where $|\mathcal{Q}|$ and $|\mathcal{P}|$ are the size of the query set and product-type set respectively.
% The feature vector can be constructed by the edge weights (Equation \ref{eq:query_vector} \& \ref{eq:pt_vector}).
% \begin{equation}
%     v_{a^i} = (w_{(a^i, b^{i}_1)}, w_{(a^i, b^{i}_2)}, ...) 
%     \label{eq:query_vector}
% \end{equation}
% \begin{equation}
%     v_{b^i} = (w_{(a^i_1, b^i)}, w_{(a^i_2, b^i)}, ...) 
%     \label{eq:pt_vector}
% \end{equation}
Based on the similarity between two nodes (e.g., the euclidean distance between their feature vectors) and a threshold $\tau$ which defines the minimal distance between two nodes to be classified into the same cluster, we conduct a round of hierarchical clustering on the query nodes first and then on the product-type nodes.
We then compute the new feature vectors of the new clusters for a next round clustering until the stop criteria are satisfied.
We keep the learned query clusters for recommendations.
Algorithm \ref{algo:iterative_clustering} summarizes our iterative clustering algorithm. 

Note that depending on whether to conduct clustering steps for product-type nodes or not, we could either only iteratively cluster (only steps \textcolor{red}{1} \& \textcolor{red}{2} in Algorithm \ref{algo:iterative_clustering}) the query nodes (marked as \textit{\textbf{IC-1}}) or apply iterative clustering on both query nodes and product-type nodes alternatively like \cite{DBLP:conf/kdd/BeefermanB00}\cite{45569} (marked as \textit{\textbf{IC-2}}).
We compare two strategies in the experiment.

\begin{figure}
    \begin{minipage}[t]{\linewidth}
    \begin{equation}
    w_{(q,p)} = \sqrt{\frac{\alpha + Clk_{(q, p)}}{\beta + Imp_{(q, p)} - Clk_{(q,p)}} (\alpha + \beta + Imp_{(q, p)} + 1)}
    \label{eq:edge_weight}
    \end{equation}
    \end{minipage}
    \begin{minipage}[t]{0.49\linewidth}
    \begin{equation}
    v_{a^i} = (w_{(a^i, b^{i}_1)}, w_{(a^i, b^{i}_2)}, ...) 
    \label{eq:query_vector}
    \end{equation}
    \end{minipage}
    \begin{minipage}[t]{0.49\linewidth}
    \begin{equation}
    v_{b^i} = (w_{(a^i_1, b^i)}, w_{(a^i_2, b^i)}, ...) 
    \label{eq:pt_vector}
    \end{equation}
    \end{minipage}
    \vspace{-8pt}
\end{figure}

\begin{algorithm}
\SetAlgoLined
\KwResult{query clusters $A_{\mathcal{Q}}$}
  initialize $A^0_{\mathcal{Q}}$, $B^0_{\mathcal{P}}$,  $\tau_{q}$ for query clustering, $\tau_{p}$ for product-type clustering, and $i = 0$ \;
%  initialization\;
 \While{True}{
  \textcolor{red}{1}. create feature vectors for each query cluster and product-type cluster by Equations \ref{eq:query_vector} \& \ref{eq:pt_vector}\;
  \textcolor{red}{2}. Apply the hierarchical clustering with $\tau_{q}$ on set $A^i_{\mathcal{Q}}$ and get new set of clusters $A^{i+1}_{\mathcal{Q}}$\;
  \textcolor{blue}{3}. \textcolor{blue}{(optional)} Apply the hierarchical clustering with $\tau_{p}$ on set $B^i_{\mathcal{P}}$ and get new set of clusters $B^{i+1}_{\mathcal{P}}$ \;
  \If{no new query cluster and product-type cluster forms}{
    $A_{\mathcal{Q}} = A^i_{\mathcal{Q}}$, return $A_{\mathcal{Q}}$;
  }
  $ i = i + 1$;
 }
 \caption{Iterative Clustering (IC)} 
 \label{algo:iterative_clustering}
\end{algorithm}

\subsection{Product Carousel Recommendations}

% todo: add a paragraph of how this model will power the product carousel recommendations, possible usecases
Only learning the query clusters for event-related aspects is not enough.
Based on the Definition \ref{def:definition}, we need to recommend products or product-types for each aspect.
To further generate the recommendations for each aspect, we leverage the click data to power the aspect-related carousels.
After all the query clusters are learned, we consider each cluster $a \in A_{\mathcal{Q}}$ as a carousel and report top $K$ largest clusters.
The clicked product-type $p$ of each cluster is ranked by its \textit{click-through} rate $\text{CTR}(a, p) = \frac{\sum_{q \in a} Clk_{(q, p)}}{\sum_{q \in a} Imp_{(q, p)}}$ in an descending order.
The top-$Z$ product-types are reported for each cluster as the recommendations for the carousel.
We summarize the entire recommender system and show how our model generate the event-based product carousel recommendations in Figure \ref{fig:workflow}.

\subsection{A Solution to \textit{Father's Day} Requests}
\begin{figure}[t]
    \centering
    \includegraphics[width=0.47\textwidth]{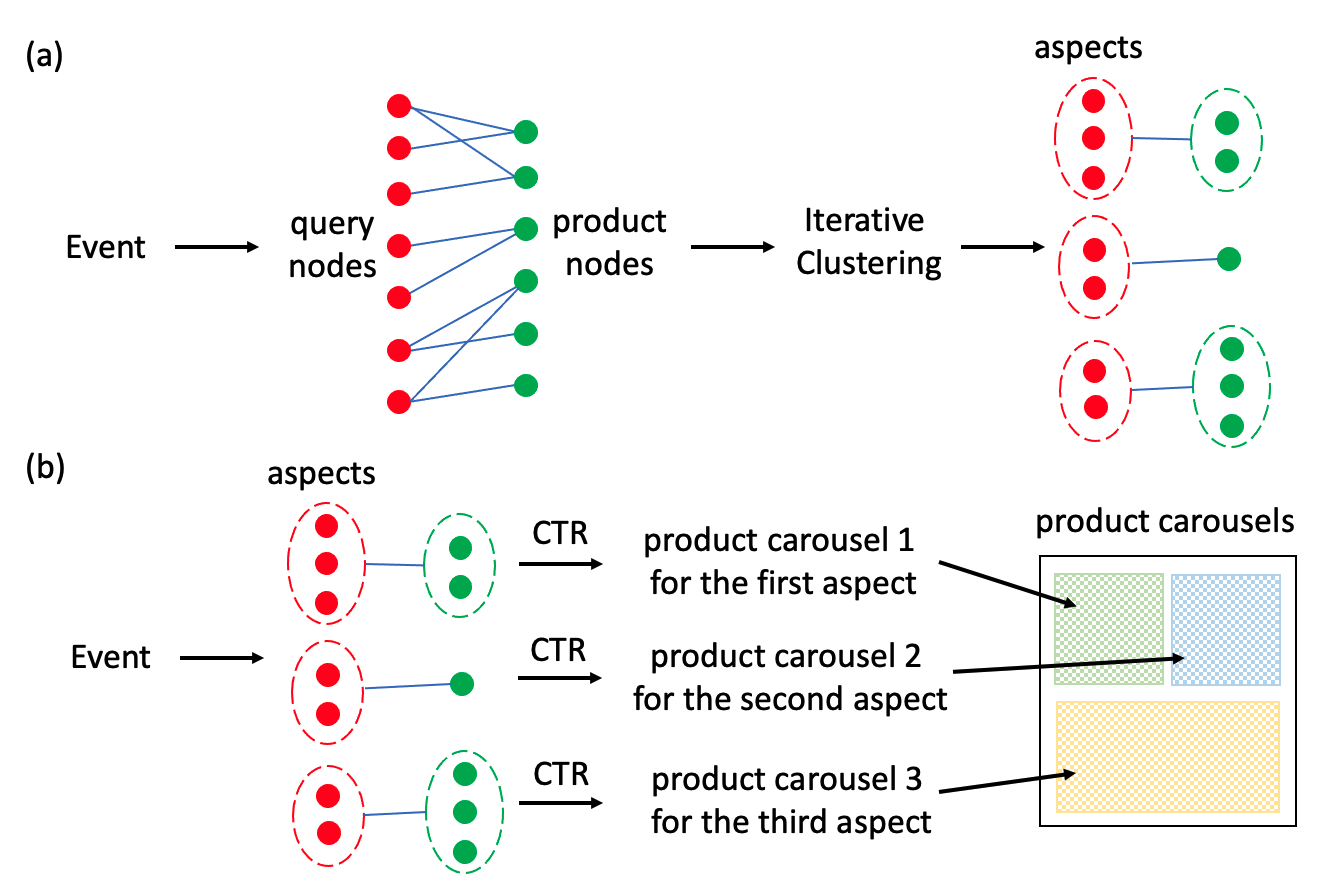}
    \caption{ (a) the training stage of our event-based product carousel recommdner system. Given an event, we create the bipartite graph and apply the iterative clustering algorithm to infer the aspects and corresponding product pools.
    (b) the recommendation stage. Given an event, our recommender system uses the inferred aspects and for each aspect, we rank the product in the corresponding product pool by the \textit{click-through} rate in descending order as the product carousel for this aspect.
    }
    \label{fig:workflow}
\end{figure}

As aforementioned, the product carousels for an event are mostly human-curated. 
Our solution can solve the request of product carousel recommendations well for an event.
For the \textit{Father's Day} example, the e-commerce platform might need to generate several related product carousels on the homepage or the post-transaction page to address the different aspect of \textit{Father's Day}.
With our model, the e-commerce platform can not only infer the different aspects of \textit{Father's Day} shopping by Figure \ref{fig:workflow}-(a), but also automatically generate the corresponding product carousel for each aspect via Figure \ref{fig:workflow}-(b). 
For another event like \textit{Super Bowl}, the e-commerce platform just needs to rerun the Figure \ref{fig:workflow}-(a) and (b) efficiently without waiting for human-curations.

\section{Experiments}
In this section, we first describe our dataset and the state-of-the-art baselines for comparison. 
After we introduce the implementation details, we present our experiments on two types of events, \textit{Valentine's Day} (a festival) and \textit{Summer Water Activity} (a seasonal activity), to show the effectiveness of our solution.

To evaluate our proposed model properly for the novel event-based product carousel recommendation problem, we try to address the crucial factors for good recommendations of product carousels for an event.
Typically, we evaluate our model by answering the following questions:
\begin{itemize}
    \item \textbf{Q1}: Does the model recommend related recommendations (\textbf{precision})?
    \item \textbf{Q2}: Do the carousels recommended by the model cover diverse aspects of an event (\textbf{heterogeneity})?
    \item \textbf{Q3}: Do the recommendations in the same carousels serve the same aspect (\textbf{cohesion})?
    \item \textbf{Q4}: Is the clustering on the product-type nodes always required for query clustering (\textbf{ablation study})?
\end{itemize}
The question of precision requests that a good model should first recommended event-related recommendations.
Next, the questions of heterogeneity and cohesion request that the model should also learn the distinct and diverse aspects of the event while the products in the carousel of each aspect should be cohesive. 

\begin{table}[]
    \centering
    \caption{Data Description}
    \label{tab:data_stats}
    \begin{tabular}{lccc}
    \hline 
         & \#Queries & \makecell[c]{\#Unique Clicked \\ Product-types} & \makecell[c]{\#Unique Impression \\ Product-types}  \\
         \hline 
         Valentine's Day & 693,556 & 999 & 2,528\\ 
         \makecell[c]{Summer Water \\ Activity} & 34,023 & 202 & 821\\ 
         \hline
    \end{tabular}

\end{table}

\begin{table}[t]
    \begin{minipage}[t]{\linewidth} 
  \centering
    \footnotesize
  \captionof{table}{Precision of Recommendations for Valentine's Day}
  \label{tab:precision}
  \begin{tabular}{cadcce}
    \hline
    $\tau$ & \textit{\textbf{IC-1}} & \textit{\textbf{IC-2}} & \textit{\textbf{HC}} & \textit{\textbf{DBSCAN}}  \\
    \hline
    $\tau = 0.1$ &  \textbf{71.43}\%	& \textbf{71.43}\% & 35.71\% & 64.29\%  \\
    $\tau = 0.2$ & \textbf{71.43}\% & \textbf{71.43}\% & 35.71\% & 57.14\% \\
    $\tau = 0.5$ & \textbf{71.43}\% & 64.29\% & 50.00\% & 35.71\% \\
  \hline
  \end{tabular}
  \end{minipage}
  
  \vspace{10pt}
  
  \begin{minipage}[t]{\linewidth}
  \centering
    \footnotesize
  \captionof{table}{Heterogeneity of Carousels for Valentine's Day }
  \label{tab:heterogeneity}
  \begin{tabular}{cadcc}
    \hline
    $\tau$ & \textit{\textbf{IC-1}} & \textit{\textbf{IC-2}} & \textit{\textbf{HC}} & \textit{\textbf{DBSCAN}} \\
    \hline
    $\tau = 0.1$ & 0.882 & \textbf{0.884} & 0.824 & 0.862 \\
    $\tau = 0.2$ &  0.879 & 0.879 & 0.844 & \textbf{0.885}  \\ 
    $\tau = 0.5$ & 0.848 & 0.860 & 0.870 & \textbf{0.895} \\
  \hline
  \end{tabular}
  \end{minipage}
  
  \vspace{10pt}
  
  \begin{minipage}[t]{\linewidth}
  \centering
    \footnotesize{
  \captionof{table}{Cohesion of Carousels for Valentine's Day}
  \label{tab:cohesion}
  \begin{tabular}{cadcc}
    \hline
    $\tau$ & \textit{\textbf{IC-1}} & \textit{\textbf{IC-2}} & \textit{\textbf{HC}} & \textit{\textbf{DBSCAN}} \\
    \hline
    $\tau = 0.1$ & 0.426 & 0.435 & \textbf{0.540} & 0.513  \\
    $\tau = 0.2$ & \textbf{0.435} & \textbf{0.435} & 0.417 & 0.377 \\
    $\tau = 0.5$ & 0.476 & \textbf{0.500} & 0.476 & 0.392  \\
  \hline
  \end{tabular}
  }
  \end{minipage}
  
  \vspace{10pt}
  
   \begin{minipage}[t]{\linewidth} 
  \centering
    \footnotesize
  \captionof{table}{Precision of Recommendations for Summer Water Activity}
  \label{tab:precision_2}
  \begin{tabular}{cadcc}
    \hline
    $\tau$ & \textit{\textbf{IC-1}} & \textit{\textbf{IC-2}} & \textit{\textbf{HC}} & \textit{\textbf{DBSCAN}} \\
    \hline
    $\tau = 0.1$ &  \textbf{32.50}\% & \textbf{32.50}\% & 30.00\% & 32.50\%  \\
    $\tau = 0.2$ & \textbf{42.50}\%  & 40.00\%  & 37.50\% & 35.00\% \\
    $\tau = 0.5$ & \textbf{47.50}\% & \textbf{47.50}\% & 45.00\% & 30.00\% \\
  \hline
  \end{tabular}
  \end{minipage}
  
  \vspace{10pt}
  
  \begin{minipage}[t]{\linewidth}
  \centering
    \footnotesize
  \captionof{table}{Heterogeneity of Carousels for Summer Water Activity }
  \label{tab:heterogeneity_2}
  \begin{tabular}{cadcc}
    \hline
    $\tau$ & \textit{\textbf{IC-1}} & \textit{\textbf{IC-2}} & \textit{\textbf{HC}} & \textit{\textbf{DBSCAN}} \\
    \hline
    $\tau = 0.1$ & \textbf{0.845} & \textbf{0.845} & 0.762 & 0.820 \\
    $\tau = 0.2$ &  0.836 & 0.831 & 0.803 & \textbf{0.843}  \\ 
    $\tau = 0.5$ & 0.845 & 0.842 & 0.840 & \textbf{0.850}  \\
  \hline
  \end{tabular}
  \end{minipage}
  
  \vspace{10pt}
  
  \begin{minipage}[t]{\linewidth}
  \centering
    \footnotesize{
  \captionof{table}{Cohesion of Carousels for Summer Water Activity}
  \label{tab:cohesion_2}
  \begin{tabular}{cadcc}
    \hline
    $\tau$ & \textit{\textbf{IC-1}} & \textit{\textbf{IC-2}} & \textit{\textbf{HC}} & \textit{\textbf{DBSCAN}} \\
    \hline
    $\tau = 0.1$ & \textbf{0.714} & \textbf{ 0.714} & \textbf{0.714} & 0.682  \\
    $\tau = 0.2$ & 0.483 & 0.500 & \textbf{0.535} & \textbf{0.535} \\
    $\tau = 0.5$ & 0.441 & 0.441 & \textbf{0.468} & 0.454  \\
  \hline
  \end{tabular}
  }
  \end{minipage}
  
\end{table}

\label{sec:experiment}
\subsection{Experimental Settings}
\noindent \textbf{Dataset}:
We collect a sample of the search query and the impression/click data on Walmart.com.
To alleviate the impact of the COVID-19 pandemic on events and also  keep the recency of the data, we collect the data for \textit{Valentine's Day} and \textit{Summer Water Activity} differently.
For \textit{Valentine's Day}, we collect the queries and the impression/click data between 2020-01-31 and 2020-02-14 (before the COVID-19 pandemic) for Valentine's Day 2020.
For \textit{Summer Water Activity}, we collect the queries and the impression/click data between 2021-06-01 and 2020-08-31 for \textit{Summer Water Activity} 2021 when the COVID-19 crisis becomes easing \footnote{https://covid19.healthdata.org/united-states-of-america?view=daily-deaths\&tab=trend}.
Table \ref{tab:data_stats} summarizes the statistics of two datasets.

% It contains $74603$ distinct queries with the keywords of  \textit{Valentine's Day} mentioned in Section 3.2 and $2528$ unique product-types.

To get the ground truth labels for the evaluation of the model precision, we asked human experts to manually identify $14$ product-types to specifically pertain to \textit{Valentine's Day} from $2528$ unique impression product-types, and $10$ product-types for \textit{Summer Water Activity} from $821$ impression product-types.

\noindent \textbf{Baselines}:
As aforementioned, we focus on the learning of multiple aspects of the target as well as the aspect-related product carousels for recommendations which is not addressed by many recommender systems.
Although frequency-based baselines such as popular items during the period of the target event might achieve considerable coverage, they couldn't be adapted to aspect learning.
Similarly, most of the product-to-product recommender systems and user-to-product recommender systems cannot address the aspect learning due to their incapability of modeling aspects of the target event.
Hence, we select the baseline models which can learn mine the aspects of the target events and can be converted to recommender systems with the same setting of our model.
To compare with the proposed models by Algorithm \ref{algo:iterative_clustering} (\textit{\textbf{IC-1}} and \textit{\textbf{IC-2}}), we consider following baseline clustering algorithms which don't require the predefined number of clusters:
\begin{itemize}
    \item Hierarchical clustering (\textbf{\textit{HC}}): the base clustering method used by our \textit{\textbf{IC}} method. 
    \item \textbf{\textit{DBSCAN}}: a density-based spatial clustering algorithm which is first proposed in \cite{DBLP:conf/kdd/EsterKSX96}.
\end{itemize}

\noindent \textbf{Implementation}:
When generating the feature vectors, we initialize $\alpha$ by the average \textit{click-through} rate of product-types in the entire dataset and set $\beta = 1 - \alpha$.
For hierarchical clustering, we use the average of the distances of each observation of the two clusters as the linkage criterion.
For both \textit{\textbf{HC}} and \textbf{\textit{DBSCAN}}, we use the euclidean distance between feature vectors for the similarity and use the same distance threshold $\tau$.
In particular, we let $\tau_q = \tau_p = \tau$ in  \textbf{\textit{IC-1}} and \textbf{\textit{IC-2}} model.
For \textit{\textbf{DBSCAN}}, we set the number of samples in a neighborhood for a point to be considered as a core point to be $3$.
For all models, we report the top 20 clusters and each cluster contains top 5 product-types.
We conduct parameter analyses on the hyperparameters $\tau$ regarding the precision, the heterogeneity and the cohesion of the carousel recommendations.

\noindent \textbf{Evaluation Metrics}: 
(1) To measure precision, we consider the percentage of ground truth product-types of the event (labeled by human experts) covered by all product-types of the model-generated product carousel recommendations. 
(2) To evaluate the heterogeneity of the carousel recommendations properly under the product recommendation scenario, we develop our measurement of the heterogeneity for each model's recommendations  $D_{model} = 1 - \left( \frac{1}{{{|A_{\mathcal{Q}}|}\choose{2}}} \sum_{a^i_{pt} \neq a^j_{pt}}J(a^{i}_{pt}, a^j_{pt}) \right)$ where $a^i_{pt}$ and $a^j_{pt}$ are two sets of product-types for the carousels $a^i$ and $a^j$ in $A_{\mathcal{Q}}$ respectively and 
$J(X, Y) = \frac{|X \cap Y|}{|X \cup Y|}$ is the Jaccard Similarity.
A higher heterogeneity score indicates a more diverse set of carousels.
(3) To also quantify the cohesion of recommendations in the same carousel properly under the product recommendation scenario, we consider the cohesion score $C_{model} = \frac{1}{S_{model}}$ where $S_{model}$ is the average number of departments cover by each carousel's recommendations. 
A less $S_{model}$ indicates more concentrated recommendations which leads to a higher cohesion score $C_{model}$.

\subsection{Results}
In this section, we address the four aforementioned questions
and report the results of the \textbf{precision}, the \textbf{heterogeneity} and the \textbf{cohesion} in Table \ref{tab:precision}, \ref{tab:heterogeneity} and \ref{tab:cohesion} for \textit{Valentine's Day} and Table \ref{tab:precision_2}, \ref{tab:heterogeneity_2} and \ref{tab:cohesion_2} for \textit{Summer Water Activity} respectively.
We highlight the globally best performance in bold with a given distance threshold $\tau$.

\noindent\textbf{Q1. Precision}: Compared with the all baseline models, our iterative clustering models (\textit{\textbf{IC-1} and \textit{\textbf{IC-2}}}) achieve the best precision score on the ground truths. 
As the distance threshold $\tau$ increases, our models present their robustness by showing stable precision scores while the performance of \textit{\textbf{HC}} and \textit{\textbf{DBSCAN}} changes along with $\tau$.
We believe that the difference of models in precision scores is because the single-round  clustering is limited to reveal similar queries which don't point to the same product-types and show less similarity based on the feature vectors. 
The iterative clustering can alleviate this situation by merging some nodes with the distance threshold $\tau$ which reduces the sparsity of the linkage between nodes and retrieves more related product-types. 
Notice that \textit{\textbf{DBSCAN}} is sensitive to the distance threshold $\tau$. 
We believe it is mainly because \textit{\textbf{DBSCAN}} directly uses the distance threshold to determine core data points and the size of their neighbors, rather than considering the average distance between clusters for clustering in \textit{\textbf{HC}}, \textit{\textbf{IC-1}} \& \textit{\textbf{IC-2}}.
Hence, the larger distance threshold leads to the less purity of clusters (the cohesion score in Table \ref{tab:cohesion}) and less related product-types in each clusters in \textit{\textbf{DBSCAN}}. 

\noindent \textbf{Q2} \& \textbf{Q3. Heterogeneity and Cohesion}:
Heterogeneity and cohesion mutually define the quality of product carousel recommendations for varied aspects of the target event.
A good product carousel recommender system should balance both the heterogeneity and the cohesion by learning diverse aspects and cohesive product recommendations within the product carousel of the aspect.
From Table \ref{tab:heterogeneity}, \ref{tab:cohesion}, \ref{tab:heterogeneity_2} and \ref{tab:cohesion_2}, we can see all four models show comparable performance.
This indicates that our iterative clustering method doesn't hurt the heterogeneity and the cohesion of the learned aspects and product carousels while increasing the coverage of the recommendations compared with the baseline clustering models.

\noindent\textbf{Q4. Ablation Study}: We also conduct the ablation study by comparing \textit{\textbf{IC-1}} and \textbf{\textit{IC-2}}.
For most cases, these two models behave similarly.
We believe their similar performance is mainly due to the imbalance between the number of query nodes and the number of product-type nodes.
Unlike search engines where a huge number of URLs are covered and clicked, an e-commerce platform usually has a limited number of product-types.
Hence, merging the clustered query nodes contributes more to the overall clustering performance than merging the clustered product-type nodes. 
We can skip the clustering steps for product-types depending on the data distribution to accelerate the overall clustering process while keeping the promising performance.

\begin{table}[t]
  \begin{minipage}[t]{\linewidth}
  \centering
    \footnotesize
  \captionof{table}{Top-5 product carousels for Valentine's Day}
  \label{tab:case_study_1}
  \begin{tabular}{cl}
    \hline
    Carousels & Top-$5$ product-types by CTR   \\
    \hline
    \makecell[c]{Carousel 1 \\ (greeting cards \& candies)} & \makecell[l]{Greeting Cards, Chocolate Candy, \\ Chocolate Assortments,  Party Bags, \\ Lollipops \& Suckers} \\
    \hline 
    \makecell[c]{Carousel 2 \\ (stuffed and plush toys)} & \makecell[l]{Stuffed Animals \& Plush Toys, \\ Gift Baskets \& Sets,Party Favors,  \\ Greeting Cards, Decoration} \\
    \hline 
    \makecell[c]{Carousel 3 \\ (women's clothing)} & \makecell[l]{T-Shirts, Blouses \& Tops,\\ Outfit Sets,  Tank Tops,  \\Sweatshirts \& Hoodies} \\
    \hline 
    \makecell[c]{Carousel 4 \\ (snacks)} & \makecell[l]{Cookies, Fruit Snacks, Cakes,  Cupcakes, \\ Snack Crackers} \\
    \hline 
    \makecell[c]{Carousel 5 \\ (nightwear \& innerwear)} & \makecell[l]{Pajamas, Nightgowns, \\ Sleepwear Robes \& Bathrobes,  \\ Pants, Lingerie Sets} \\
  \hline
  \end{tabular}
  \end{minipage}
  \vspace{5pt}
  
  \begin{minipage}[t]{\linewidth}
  \centering
    \footnotesize
  \captionof{table}{Top-5 product carousels for Super Bowl}
  \label{tab:case_study_2}
  \begin{tabular}{cl}
    \hline
    Carousels & Top-$5$ product-types by CTR   \\
    \hline
    \makecell[c]{Carousel 1 \\ (party supplies)} & \makecell[l]{Party Supply Sets, Party Favors, Tableware Plates, \\ Balloons, Party Banners} \\
    \hline 
    \makecell[c]{Carousel 2 \\ (apparel)} & \makecell[l]{T-Shirts, Hats, Sweatshirts \& Hoodies, \\  Pins \& Brooches, Blouses \& Tops} \\
    \hline 
    \makecell[c]{Carousel 3 \\ (game watching)} & \makecell[l]{Televisions, TV Shows, Projector Screens,\\ TV Antennas, Computer Monitors} \\
    \hline 
    \makecell[c]{Carousel 4 \\ (snacks)} & \makecell[l]{Snack Chips, Pretzels, Granola, Popcorns, \\ Ice Cream \& Frozen Yogurt}\\
    \hline 
    \makecell[c]{Carousel 5 \\ (decorations)} & \makecell[l]{Photographic Art, Art Prints,  Posters, \\ Outdoor Flags \& Banners, Plaques \& Signs } \\
  \hline
  \end{tabular}
  \end{minipage}
   \vspace{5pt}
  
  \begin{minipage}[t]{\linewidth}
  \centering
    \footnotesize
  \captionof{table}{Top-5 product carousels for Summer Water Activity}
  \label{tab:case_study_3}
  \begin{tabular}{cl}
    \hline
    Carousels & Top-$5$ product-types by CTR   \\
    \hline
    \makecell[c]{Carousel 1 \\ (swimming pool)} & \makecell[l]{Swimming Pools, Pool Toys \& Floats, \\Lawn Water Slides, Inflatable Bouncers \\ Pool Chemicals} \\
    \hline 
    \makecell[c]{Carousel 2 \\ (pool toys)} & \makecell[l]{Pool Toys \& Floats, Life Jackets \& Vests, \\ Gag Toys, Bath Toys, Water Guns} \\
    \hline 
    \makecell[c]{Carousel 3 \\ (swimming suits)} & \makecell[l]{Swimwear Bottoms, Swimsuit Sets, \\Athletic Rash Guards, One-Piece Swimsuits, \\ Bikini \& Tankini Tops} \\
    \hline 
    \makecell[c]{Carousel 4 \\ (pool maintenance)} & \makecell[l]{Pool Chemicals, Pool Skimmers, \\ Pool \& Pond Test Kits, Pool Chemical Dispensers} \\
    \hline 
    \makecell[c]{Carousel 5 \\ (protection gears)} & \makecell[l]{Sport Goggles, Hats, Work Safety Eye Protection, \\ Swimming Nose Clips, Snorkels} \\
  \hline
  \end{tabular}
  \vspace{-10pt}
  \end{minipage}
\end{table}

\subsection{Case Study}
We visualize top 5 clusters inferred by our iterative clustering algorithm \textit{\textbf{IC-2}} as 5 carousels for \textit{Valentine's Day}, \textit{Super Bowl} and \textit{Summer Water Activity} in Table \ref{tab:case_study_1}-\ref{tab:case_study_3} respectively, where the top 5 product-types of each carousel are listed to show the recommendations.
We can see that meaningful carousels are learned.
For \textit{Valentine's Day}, our model not only mines the classic aspect of cards and candies, but also proposes many well-fitting aspects such as stuffed toys, nightwear for gifts.
For \textit{Super Bowl}, our model also proposes game watching and decoration aspects in addition to the classic aspect of parties and snacks for the super bowl event.
For \textit{Summer Water Activity}, our model can even help the customers to remind the aspect of pool maintenance and protection gears while shopping the aspect of pools, toys and swimming suites.

\section{Conclusion}
In this paper we formally define the novel event-based product carousel recommendation problem and propose an effective recommender system via iterative clustering over the query-click bipartite graph.
With extensive experiments and the case study covering different types of events, our recommender system shows meaningful product carousels for the target event as recommendations with good precision

\bibliographystyle{IEEEtran}
\bibliography{main}

\end{document}